\documentclass{emulateapj}

\usepackage{amsmath} 
\usepackage{amsfonts} 
\usepackage{amssymb} 
\usepackage{xspace}
\usepackage{natbib}
\usepackage{wasysym}
\newcommand{\msun}{\ensuremath{M_\odot}\xspace}

\newcommand{\semimaj}{\ensuremath{\it{a}}\xspace} 	
\newcommand{\semimajC}{\ensuremath{\it{a}}\xspace} 	
\newcommand{\eccC}{\ensuremath{\it{e}}\xspace}

\newcommand{\massp}{\ensuremath{M_{\mathrm{p}}}\xspace} 	
\newcommand{\mjup}{\ensuremath{M_{\mathrm{J}}}\xspace}

\newcommand{\mearth}{\ensuremath{M_{\oplus}}\xspace}

\newcommand{\gcm}{\ensuremath{\rm{g\,cm^{-3}}}\xspace}

\newcommand{\wmf}{\ensuremath{\rm{WMF}}\xspace} 	
\newcommand{\water}{\ensuremath{\rm{H_{2}O}}\xspace} 	
\newcommand{\logwmf}{\ensuremath{\rm{log(WMF)}}\xspace} 	

\newcommand{\mocean}{\ensuremath{\it{M}_{\rm{ocean}}}\xspace} 	
\newcommand{\mwater}{\ensuremath{\it{M}_{\rm{\water}}}\xspace} 	
\newcommand{\massf}{\ensuremath{\it{M}_{\rm{f}}}\xspace}

\newcommand{\semimajin}{\ensuremath{\it{a}_{in}}\xspace} 	

\begin{document} 

\title{The Effect of Planets Beyond the Ice Line on the Accretion of Volatiles by Habitable-Zone Rocky Planets} 
  \author{Elisa V. Quintana}\affil{SETI Institute, 189 Bernardo Avenue \#100  Mountain View, CA 94043 \\ Space Science and Astrobiology Division 245-3, NASA Ames Research Center, Moffett Field, CA 94035}  
  \email{elisa.quintana@nasa.gov} 
  \author{Jack J. Lissauer}\affil{Space Science and Astrobiology Division 245-3, NASA Ames Research Center, Moffett Field, CA 94035}

\begin{abstract}

Models of planet formation have shown that giant planets have a large impact on the number, masses and orbits of terrestrial planets that form.  In addition, they play an important role in delivering volatiles from material that formed exterior to the snow-line (the region in the disk beyond which water ice can condense) to the inner region of the disk where terrestrial planets can maintain liquid water on their surfaces.  We present simulations of the late stages of terrestrial planet formation from a disk of protoplanets around a solar-type star, and we include a massive planet (from 1 \mearth to 1 \mjup) in Jupiter's orbit at $\sim$5.2 AU in all but one set of simulations.  Two initial disk models are examined with the same mass distribution and total initial water content, but with different distributions of water content. We compare the accretion rates and final water mass fraction of the planets that form. Remarkably, all of the planets that formed in our simulations without giant planets were water-rich, showing that giant planet companions are $\it{not}$ required to deliver volatiles to terrestrial planets in the habitable zone.  In contrast, an outer planet at least several times the mass of Earth may be needed to clear distant regions from debris truncating the epoch of frequent large impacts. Observations of exoplanets from radial velocity surveys suggest that outer Jupiter-like planets may be scarce, therefore the results presented here suggest the number of habitable planets that reside in our galaxy may be more than previously thought.

\end{abstract}
\clearpage 

\section{Introduction}

Oceans cover more than two-thirds of the surface of our planet, and water and other volatile compounds are the dominant constituents of living organisms on Earth.  Yet by cosmic standards, Earth is highly deficient in volatiles.  The condensed component of a solar composition mixture that is cool enough for all of the H$_2$O to be in solid form is $\gtrsim$ 50\% ice by mass.  In contrast, the Earth's oceans and other near-surface reservoirs represent only 0.03\% of our planet's mass, with several times this amount of water thought to lie in the mantle \citep{Marty.Yokochi:2006}.  Nonetheless, the Earth was able to accrete enough water and other volatile constituents such as carbon and nitrogen to support life as we know it.  Understanding where these volatiles originated from, and how they made their way to Earth, is important in determining the likelihood that life exists beyond our Solar System.

The region of the disk where Earth now resides was too hot for water ice to have condensed during Earth's formation, therefore the bulk of Earth's water must have originated from other reservoirs.  The leading theories for the origin of Earth's water have focused on icy comets and water-rich asteroids as the main sources.  Comparisons of the isotopic deuterium to hydrogen (D/H) ratio measured in Earth's oceans, atmospheres and present-day mantle \citep{Lecuyer.etal:1998} to the D/H ratio measured in meteorites and comet spectra have provided valuable clues used to constrain these theories.

The contribution of water from a bombardment of comets is thought to be limited to $\lesssim$10\% of the Earth's crustal water \citep{Morbidelli.etal:2000, Morbidelli.etal:2012,Marty:2012}.  These limitations were partially based on the D/H ratios measured from several comet samples that were found to be more than double the D/H ratio of water found on Earth \citep{Balsiger.etal:1995, Bockel.etal:1998, Eberhardt.etal:1995, Meier.etal:1998}.  The recent discovery of comet Hartley 2 with an Earth-like D/H ratio \citep{Hartogh.etal:2011} sparked new interest into comets as a viable source of Earth's water.  However, \citet{Alexander.etal:2012} points out that Earth would have accreted entire comets, not just cometary ice, which have compositions that are more deuterium-rich than water due to the large amounts of organic material.  Despite these developments, a more serious problem that remains is that the collision probability of comets with Earth is very low \citep{Morbidelli.etal:2000}, most likely too low to provide the bulk of water collected by Earth.

The best explanation for the origin of Earth's water is that water-rich chondritic planetary embryos formed in the outer asteroid region and were perturbed inwards during Earth's formation \citep{Morbidelli.etal:2000}. Geochemical data from meteorite samples have uncovered a rough correlation between water content and the heliocentric distance in the disk from which they are thought to have originated.  Enstatite and ordinary chondrites from the inner asteroid belt ($\sim$1.8 -- 2.8 AU) tend to be dry \citep{Fornasier.etal:2008, Binzel.etal:1996}, whereas carbonaceous chondrites from the middle and outer asteroid belt (approximately 2.5 -- 4 AU) are relatively water rich (up to $\sim$20\% in mass) \citep{Mason:1963}.  The D/H ratio in Earth's oceans is nearly identical to the mean D/H ratio found in carbonaceous chondrites \citep{Dauphas.etal:2000}, supporting the outer asteroid region as the primary source of Earth's water.  In addition, the relative abundances of hydrogen, carbon, and noble gases in Earth have been found to be roughly chondritic \citep{Marty:2012}.  For our dynamical analysis herein, we  adopt the model that chondritic material from the outer asteroid region is the principal source of Earth's volatiles.

Giant planets, which likely form prior to the epoch of terrestrial planet formation that we model herein \citep{Lissauer:1987, Lissauer.etal:2009,Movshovitz.etal:2010}, have long been thought to be a crucial factor in inducing the radial mixing among growing protoplanets that is needed for water delivery.  Numerous simulations of the accretion of planetesimals from volatile-rich regions of protoplanetary disks by terrestrial planets have been performed \citep{Raymond.etal:2004,Raymond.etal:2005a,Raymond.etal:2005b,Raymond.etal:2006a,Raymond.etal:2006b,Raymond.etal:2007,Raymond.etal:2007b,Raymond:2006}.  These studies included various different disk surface density profiles, star masses, and the (in many cases large) influences that varying these parameters have on water content of the planets that are ultimately formed.  Giant planets exterior to the terrestrial region around low mass stars were found to reduce accretion timescales, leading to typically drier terrestrial planets \citep{Raymond.etal:2005a}.  High resolution (large initial number of planetesimals) simulations found that giant planet eccentricities have a large influence on the amount of water delivered to terrestrial planets \citep{Obrien.etal:2006, Raymond.etal:2009}.  Simulations that include a solar-type stellar companion show similar results \citep{Haghighipour.etal:2007}, as expected, although simulations of accretion in binary star systems are limited due to the much larger parameter space (stellar masses and orbits) available.   

We investigate herein the effects of different-sized planets in Jupiter's orbit on the accretion of volatiles by terrestrial planets in or near the habitable zone of a Sun-like star during the late stages of planet formation.  We first study the case of planet formation around an isolated star with no distant companions perturbing the system.  We then perform analogous simulations but include a planet of mass 1 \mearth, 10 \mearth or 1 \mjup on an initial orbit comparable to Jupiter's current orbit (semimajor axis \semimajC = 5.2 AU and \eccC = 0.05).   We examine two additional variations of systems that include a Jupiter-like planet: Jupiter on a circular orbit (\eccC = 0) and Jupiter with \eccC = 0.05 with the addition of a Saturn-like planet (\semimajC = 9.5 AU and \eccC = 0.05).

Our $N$-body simulations begin at the epoch of planet formation in which planetesimals and planetary embryos have already formed in a disk, as have any giant planets included, and the gas in the disk has been dispersed.  We examine two different models for the water distribution in the disk and follow the accretion evolution of the bodies for 1 Gyr.  Our approach employs moderate-resolution simulations that have sufficiently modest computational requirements to allow us to perform enough simulations to disentangle effects of the companion body from stochastic variations that are an important aspect of terrestrial planet growth.  Although these models cannot provide ab initio estimates of the water accreted by terrestrial planets, models of this type are well suited for comparing the relative amounts of water accreted by terrestrial planets with different outer planet companions perturbing the system.  Our results are presented in a manner that allows for the incorporation of any model of the distribution of volatiles within the disk, provided these volatiles don't substantially alter the mean densities of the bodies. 

The next section describes our initial conditions and numerical model.  Section 3 presents the results of our accretion simulations and the volatile inventory of the final planetary systems that form, and we summarize our results in Section 4.

\section{Model}
\citet{Chambers:2001} found that numerical simulations of the final phases of terrestrial planet growth that began with a bimodal mass distribution consisting of many Mars-sized embryos embedded in a disk of Moon-sized planetesimals yielded a configuration similar to the inner Solar System in many respects.  Our simulations begin at this epoch and we follow the evolution of embryos and planetesimals (with a 1:10 mass ratio) as they collide and form into terrestrial planets.  Outer planets, if included, are assumed to have already formed.  Our model begins with a total disk mass of 4.85 \mearth and is composed of 26 embryos, each with an initial mass of 0.0933 \mearth (2.8$\times$10$^{-7}$ \msun), and 260 planetesimals of mass 0.00933 \mearth (2.8$\times$10$^{-8}$ \msun).  All bodies are assumed to have a density of 3 \gcm.

The disk extends from 0.35 AU to 4 AU from the Sun with a surface density of solids that varies as $a^{-3/2}$.   The embryos and planetesimals each begin on nearly circular (0 $\leq$ $e$ $\leq$ 0.01) and coplanar (0$^\circ$ $\leq$ $i$ $\leq$ 0.5$^\circ$) orbits, and the angular orbital elements (argument of periastron, longitude of ascending node and mean anomaly) were selected at random between 0$^\circ$--360$^\circ$.  The same random set was used for all simulations apart from minuscule variations described below.

Simulations have been performed using a similar bimodal disk around the Sun \citep{Chambers:2001, Quintana.etal:2002} and in binary star systems \citep{Quintana.etal:2002,Quintana.etal:2006,Quintana.etal:2007,Quintana.etal:2010}, although the disks in those studies contained a smaller number of embryos (14) and planetesimals (140) and extended out to only 2 AU from the central star or stars.  By extrapolating the disk beyond the snow line (the region in the disk where temperatures are low enough for ice condensation, which is near $\sim$2.5 AU around a solar-type star such as that used in our simulations), we can examine the dynamics and radial mixing of potentially volatile-rich planetesimals and embryos formed beyond this snow line. 

The initial water content in our standard disk was chosen to match that used in simulations performed by \citet{Raymond.etal:2004}.  A step-distribution was used in which bodies inside of 2 AU are given a water mass fraction (\wmf) of 0.001\%, those with 2 AU -- 2.5 AU are given \wmf = 0.1\%, and those beyond 2.5 AU have \wmf = 5\%.  Although the distribution of  \water and volatiles in a protoplanetary disk depends on numerous factors (including properties of the molecular cloud from which the star and disk system formed, along with masses, orbits and formation timescales of bodies in the disk and of any stellar or planetary companions), this step disk distribution was chosen as an approximation based on meteorite data as described in Section 1.  To quantify the water that the planets accrete, we define a unit of 1 \mocean  = 1.5$\times$10$^{24}$ g = 2.5$\times$10$^{-4}$ \mearth to be an approximation of the mass of water in all of Earth's oceans.  The amount of \water below the Earth's surface is not well constrained, but typical estimates are several \mocean \citep{Marty:2012}. The total amount of water in our initial disk is $\sim$340 \mocean.

We also examine a second disk model that has the same mass distribution and total water content, but the water is distributed among the bodies in the disk using a smooth power-law function.  To compute this function, we kept the water content at the inner edge of the disk the same (\wmf = 0.001\% for the innermost planetesimal) and then solved for the exponent $p$ in the following function that would keep the total amount of water equal to that contained in the step disk model:
\begin{eqnarray}	
\rm{\wmf}(\semimaj) = \rm{\wmf}(\semimajin) \times \left(\frac{\semimaj}{\semimajin}\right)^p,
\end{eqnarray}
where \semimajin is the semimajor axis of the innermost body, $a$ is the semimajor axis for a given embryo/planetesimal, and $p$ = 3.741.  Note that we neglect any effects of \water on the densities in order to facilitate the scaling of results.  Alternative disk water distributions can then be examined  without having to run additional accretion simulations. 

Figure 1 shows the distribution of mass and water for both disk models (which we refer to as `step disk' or `power-law disk').  Here, and in subsequent figures, the symbol size of the embryos and planetesimals are proportional to their physical size.  The symbol colors represent their fraction of water by mass on a scale where red bodies are dry (\logwmf = $-$5) and blue bodies are water-rich (\logwmf = $-$1.3). The top left panels of Figures 2 and 3 (described in the following section) show a view of the initial water distribution in the $a$-$e$ plane for both disk models.

We use the $\it{Mercury}$ integration package to follow the evolution of all bodies in the disk subject to mutual gravitational interactions and inelastic collisions (perfect accretion) for 1 Gyr.  The collisional history is tracked by recording the mass and orbital parameters involved in each impact. The larger body in a collision is taken to be the target, which `grows' by collecting the mass of the impactor, and the mass and orbit of the impactor are no longer recorded after the impact time.  We keep track of the semimajor axis (which provides information on the water content) of all impactors that a final planet accretes.  If a collision involves two bodies of the same mass (such as two planetesimals), the body with the smaller semimajor axis is taken to be the target. 

To account for the stochastic nature of $N$-body systems with close encounters, 5 or 6 simulations were performed for each configuration with very small changes in the initial conditions: a single planetesimal near 1 AU was displaced by 1 -- 5 meters along its orbit prior to the integration.  In all of our simulations, the mass of the primary star was 1 \msun, and the mass of the planetary companion at 5.2 AU, if included, ranged from 1 \mearth to 1 \mjup.  We present our results in the next section.

\section{From Planetary Embryos to Planets}
The accretion evolution of a simulation with Jupiter and Saturn, both on initial orbits having $e$ = 0.05, is shown in Figures 2 and 3.  The only difference between these figures is the initial \water distribution in the disk (step or power-law disk).  Each panel shows the semimajor axes and eccentricities of the bodies in the disk for a given integration time.  This simulation, which formed three planets within 1.5 AU along with a single outer planetesimal, was chosen from the set of Sun-Jupiter-Saturn runs because it produced the most Earth-like planet in terms of mass (0.9 \mearth) and orbit (1 AU).  Herein, we define a `planet' as a body that grew at least as massive as the planet Mercury ($\sim$0.06 \mearth), which includes any single embryo or a body consisting of at least 7 planetesimals. Note that planets can be smaller or less massive than this \citep{Barclay.etal:2013}, but we keep this prescription for simplicity and comparison with our own Solar System.  

The planet that formed near 1 AU accreted 39 \mocean of water from the step disk (Figure 2) and 19 \mocean from the power-law disk (Figure 3).  The planets on either side of this, however, were very dry ($<$ 0.1 \mocean) when formed from the step disk, and only moderately wet ($<$ 4 \mocean) in the power-law disk case.

A simulation of the step disk around the Sun, with no outer planets perturbing the system, is shown in Figure 4.  Note the semimajor axis scale is nearly twice that shown in Figures 2 and 3.   Eight planets formed between 0.5 -- 7 AU from the Sun, and an additional 47 planetesimals remained in this system out to 28 AU.  Five of the planets accreted over 10 \mocean and three formed moderately wet with several \mocean.  Similar water contents were found in the final planets when the step disk was replaced with the power-law water distribution (not shown).
 
The inclusion of outer giant planets perturbs more mass out of the system and reduces accretion timescales.  As shown in Figures 2 -- 4, the final planets are formed within 100 -- 200 Myr when Jupiter/Saturn are included, whereas bodies are still actively accreting throughout the 1 Gyr Sun-only simulations.  These results are consistent with previous simulations of terrestrial planet formation \citep{Chambers:2001, Quintana.etal:2006, Quintana.etal:2002}. It is clear that giant planets have a large influence on the dynamics of terrestrial planet formation.  Their influence on water delivery to the terrestrial region is discussed further below.

\subsection{Final Planetary Systems}
Figures 5 -- 8 present the final systems that formed in simulations that included Jupiter and Saturn (Figure 5), a 10 \mearth planet (Figure 6), a 1 \mearth planet (Figure 7) and around an isolated star (Figure 8).  Note the expanded range in semimajor axis shown in Figures 6 -- 8 ($a$ $<$ 10 AU) compared to that of Figure 5 ($a$ $<$ 5 AU).  To account for the stochastic nature of these $N$-body systems, 5 or 6 integrations were performed for each configuration, keeping all initial conditions the same except for a 1 -- 5 meter shift of one planetesimal prior to the integration.  The only difference between the left and right halves of the figures is the initial distribution of \water in the disk: panels in the left columns show results from a step disk \wmf distribution, and those in the right columns show systems that formed from the smooth power-law \wmf disk.  The colors and symbols are as described in Figure 2, and the relative size of Earth and Mercury are shown in the legend for comparison (here, Earth is given a conservative \water estimate of 10 \mocean).  The final systems formed from the two additional sets of simulations that included Jupiter (either on an $e$ = 0 or $e$ = 0.05 initial orbit) without Saturn are not shown, but the results are presented in Tables 1 and 2 and discussed in this section.

The first thing to notice is the similarity of the final planetary systems that formed $\it{within \; 2 \;AU}$ of the Sun in all of the simulations.  From 2 -- 4 planets formed, and at most 1 planetesimal remained, within 2 AU for all of these simulations.  Beyond 2 AU, differences in outer planet configurations have a strong influence on the disk.  To examine these differences among each type of system more quantitatively, we present the following statistics in Table \ref{tbl-1}:

$\bullet$ The number of planets, $N_p$, that are at least as massive as the planet Mercury (0.06 \mearth) with \semimaj $<$ 2 AU (column 2) and \semimaj $>$ 2 AU (column 6).  

$\bullet$ The number of minor planets, $N_m$, with \massp $<$ 0.06 \mearth that remained within \semimaj $<$ 2 AU (column 3) and beyond 2 AU (column 7).

$\bullet$ The percentage of the initial disk mass that composes the final planets, $m_p$, that formed within 2 AU (column 4) and beyond 2 AU (column 8).

$\bullet$ A radial mixing statistic, $S_r$, given by
\begin{equation}
S_r=\left[\sum_j \frac{m_j|a_{i,j}-a_{f,j}|}{a_{f,j}}
  \right]/\sum_j m_j,
\end{equation}
where $a_{i}$ and $m$ are the initial semimajor axis and mass of each embryo and planetesimal that becomes incorporated into a final planet, and $a_{f}$ is the semimajor axis of that planet. This is essentially a measure of the degree of radial mixing of material in the disk.

$\bullet$ The percentage of the initial disk mass that was lost from the system by colliding with the Sun, $m_{l_\star}$ (column 9).

$\bullet$ The percentage of the initial disk mass that was ejected from the system, $m_{l_{\infty}}$ (column 10).   This value also includes the masses of (the small number of) bodies that collided with the giant planet companion.

 The simulations in Table \ref{tbl-1} are labeled `Run$X_Y$' (column 1), where $X$ describes the system (0 = Sun-only, 1 = 1 \mearth, 10 = 10 \mearth, J = Jupiter with $e$ = 0.05, J0 = Jupiter with $e$ = 0, and JS = Jupiter $+$ Saturn, both with $e$ = 0.05) and $Y$ is a given realization of that system ($a$ -- $f$).

In the initial disk (at the start of our simulations), 51\% of the mass is less than 2 AU from the Sun.  The percentage of this initial disk mass that remains in the planets that orbit within 2 AU averages $\sim$53\% in most sets of simulations, with the value reduced to 48\% with eccentric Jupiter and to 44\% in the set that includes both Jupiter and Saturn.  Beyond 2 AU, the number of bodies and the amount of mass remaining in the system differs significantly among the various configurations, as does the fate of the mass that is lost.  In simulations with the Sun-only, 2 -- 5 planets and 46 planetesimals on average remained beyond 2 AU, composed of 43\% of the initial disk mass. About 3\% of the initial disk mass was lost in these simulations via ejection from the system or collisions with the Sun.  When a 1 \mearth planet was included in Jupiter's orbit, a larger fraction of mass (10\%) was lost from the system, and fewer planetesimals remained in the system on average.  The mass distribution changes significantly when a 10 \mearth planet is introduced.  From 2 - 5 planets remain beyond 2 AU, consistent with the Sun-only and 1 \mearth planet systems.  However, a much smaller number of planetesimals remain (6 on average), primarily because a large fraction of the initial disk, 26\% on average, is ejected from the systems.  Only a few percent of the initial disk mass was perturbed into the star in the simulations described thus far.

In all simulations that included a Jupiter-like planet, only 1 planet on average and at most a single planetesimal remained beyond 2 AU.  The systems without Saturn produced fairly consistent results, less than about 10\% of the initial disk mass collided with the Sun, and 35 -- 47\% was ejected.  A larger percentage of the initial mass was lost when both Jupiter and Saturn were included,  21\% on average was perturbed into the Sun and 34\% was ejected.  This is likely attributed to the strong perturbations induced by the $\nu_6$ secular resonance -- predominant near 2.1 AU -- due to the presence of Saturn.  The radial mixing statistic ($S_r$) was about 0.6 on average for the sets of simulations with either no distant perturber or one on a circular orbit.  $S_r$ was somewhat lower ($\sim$0.5) for the simulations with Jupiter only but on an eccentric orbit, and was even lower (0.4 on average) for the simulations with Jupiter and Saturn, which is not surprising considering the amount of mass that was lost from these systems.

\subsection{Characteristics of Planets Near 1 AU}
We next examine the `Earth-analogs' that formed in these simulations and discuss their chances for accreting and retaining water over the 1 Gyr simulations.   We define an Earth-analog as a planet that formed between the orbits of Venus and Mars (0.72 $< a < $ 1.52) with a mass exceeding 0.5 \mearth.  Properties of the final Earth-analog planets are given in Table \ref{tbl-3} and include the final mass, orbital elements ($a$, $e$, and $i$), the number of oceans accreted (\mwater) and \logwmf for both disk models. Table \ref{tbl-3} also provides the times of the final collisions that each Earth-analog has with an embryo ($t_{EM}$) and a planetesimal ($t_{PL}$).  

All but one of our 33 simulations produced at least one planet that is Earth-like by our definition, and none of the simulations produced more than two such planets\footnote{If the minimum mass requirement is raised to 0.8 \mearth, all but one simulation still produce an Earth analog.  However, only three simulations produced two such analogs, and in these cases one orbited near each of the boundaries of the region that we considered.}.  The eighteen simulations with a Jupiter analog produced a total of 23 such planets, whereas the other 15 simulations produced a total of 18 Earth analogs, an insignificant difference. Orbits and masses of these Earth analogs do not show any systematic dependance on the outer planet configuration.  However, there are large differences in the amount of water accreted, and even more profound variations in the duration of the accretionary epoch.  

For comparison among the sets, we consider a planet to be `water-rich' if it accreted at least 10 \mocean of water, `water-poor' (or `dry') if it accreted $<$2 \mocean, and `moderately wet' for intermediate values.  All but one of the Earth-analogs that formed in the Sun-only systems from the step disk distribution were water-rich (the other planet accreting just under 10 \mocean), and all seven of these planets that formed from the power-law disk were water-rich.  Similarly, each of the 1 \mearth systems produced a water-rich Earth-analog from both step and power-law disks.

 Systems that included Jupiter-only on an eccentric orbit and the Jupiter/Saturn set produced the only water-poor planets, and they were all from the step disk.  Five water-rich Earth-analogs formed in the eccentric Jupiter runs (in both step and power-law disk sets), and two other Earth-analogs from this system were either water-poor (from the step disk) or moderately wet (from the power-law disk, although these were on the lower end, between 2 -- 4 \mocean).  The system with Jupiter and Saturn produced nine Earth-analogs in total, and five of these from the step disk were dry, and the remaining (from both disks) were roughly split between being water-rich and moderately wet.  Changing the initial distribution of water for a given set had little effect on the average water content of the Earth-analogs that formed (Table 2), although a much larger number of simulations is probably needed to determine if there are any significant differences.

Of the 16 Earth-analogs that formed in systems with an eccentric Jupiter (with or without Saturn), 8 completed their accretion prior to 100 Myr, and accretion was complete for 5 others prior to 200 Myr.  In sharp contrast, all twelve of the Earth-analogs in systems without an outer planet more massive than Earth suffered a collision subsequent to 400 Myr.  

The simulations with Jupiter on a circular orbit and those with a 10 \mearth companion on an eccentric orbit produced results intermediate between those systems strongly perturbed  by an eccentric Jupiter and systems lacking an outer perturber more massive than Earth in terms of both water delivery and time of last impact.

\subsection{Loss of Volatiles in Collisions}

Our simulations assume perfect accretion (all collisions lead to mergers). Giant impacts, however, have the potential to strip off volatiles and atmospheres from potentially habitable planets.  Consequences of collisions are diverse and can fall into several categories (merging, net growth with fragmentation, net erosion, etc.) depending on the velocities, impact angles and mass ratios of the target and impactor \citep{Agnor.etal:2004, Asphaug.etal:2006, Asphaug:2010}.  Collision outcomes also depend on more detailed characteristics of the bodies, such as composition \citep{Leinhardt.etal:2012,Stewart.Leinhardt:2012}. Including fragmentation in $N$-body algorithms is possible, but following the evolution of fragmented material (whether it is re-accreted, dispersed into space or falls into the central star) becomes computationally expensive, and most ejecta that are not vaporized or pulverized into very small particles will be accreted by planets \citep{Chambers:2013}.  More studies are needed to realistically examine the effects of giant impacts on habitability throughout the formation of terrestrial planets.

\section{Discussion} 
We have performed 33 simulations of the late stage of planet formation with and without outer planet companions perturbing the disk. Although giant planets can have a profound effect on the types of planetary systems that form, we found that they are $\it{not}$ required to provide the radial mixing needed for volatile material from beyond the snow line to accrete onto terrestrial planets in the habitable zone.  We present two water mass fraction disk models, but we provide in the electronic edition of this article a full table of all final planets and their composite embryos and planetesimals (and their initial semimajor axes) to allow for the incorporation of any distribution of \water in the initial disk.  Table \ref{tbl-2} provides a sample of the electronic table.  Results can also be scaled for different stellar types with the formulae presented in \citet{Quintana.etal:2006}.

Can we conclude that most terrestrial planets formed around isolated stars are likely habitable? Not quite -- a more serious problem for the habitability of terrestrial planets in systems lacking giant planets is that small bodies persist beyond 2 AU for far longer.  This allows the tail of the accretionary epoch to extend well beyond that within our Solar System.  Impacts of objects an order of magnitude less massive than the planetesimals in our simulations (and therefore probably much more numerous in a realistic protoplanetary disk) are still so large that their accretion onto an Earth-like planet would produce environmental damage probably sufficient to wipe out all life as we know it \citep{Zahnle.Sleep:1997}.  Thus, without giant planets, devastating impacts might well persist for billions of years, rendering Earth-like planets unsuitable for all but perhaps the simplest and most rapidly formed life.

\acknowledgements
This work was funded in part by the NASA Ames Team of the NASA Astrobiology Institute.  E.V.Q. thanks Tom Barclay for useful discussions and assistance with the figures.

\clearpage
\begin{deluxetable}{lccccccccc}
%\rotate
\tablecolumns{10} 
\tabletypesize{\scriptsize} 
\tablecaption{Statistics for Final Planetary Systems. 
\label{tbl-1}} 
\tablewidth{0pt} 
\tablehead{
\colhead{$\bf \rm{Run}$} &
\colhead{$\bf N_p $}   &
\colhead{$\bf N_m$}   &
\colhead{$\bf m_p $} &
\colhead{$\bf S_r$} &
\colhead{$\bf N_p $}   &
\colhead{$\bf N_m$}   &
\colhead{$\bf m_p $} &
\colhead{$\bf m_{l_\star}$} &
\colhead{$\bf m_{l_\infty}$}  \\
\colhead{} & 				%Run
\colhead{ ($a$$<$2 AU)} & 	%Np
\colhead{ ($a$$<$2 AU) } & 	%Nm
\colhead{ ($a$$<$2 AU, \%) } &	%mp
\colhead{} &				%Sr
\colhead{ ($a$$>$2 AU) } & 	%Np
\colhead{ ($a$$>$2 AU) } & 	%Nm
\colhead{ ($a$$>$2 AU, \%) } & 	%mp
\colhead{ (\%)} & 		%mlstar
\colhead{ (\%)} }
\startdata 		
    & & & & & & & & &  \\
 Run0$_{a}$ & 3 & 0 & 54.6 &   0.67 & 2 & 50 & 43.4 & 1.0 & 1.0 \\ 
 Run0$_{b}$ & 3 & 0 & 53.7 &   0.75 & 5 & 47 & 42.1 & 2.7 & 1.5  \\ 
 Run0$_{c}$ & 3 & 1 & 60.2 &   0.56 & 3 & 52 & 37.5 & 0.8 & 1.5  \\ 
 Run0$_{d}$ & 3 & 0 & 39.8 &   0.53 & 4 & 45 & 56.5 & 2.7 & 1.0  \\ 
 Run0$_{e}$ & 3 & 0 & 59.2 &   0.58 & 4 & 38 & 37.3 & 1.6 & 1.9  \\ 
 $\bf Run0_{ave} $ & $\bf 3.0$ & $\bf 0.2$ & $\bf 53.5$ &  $\bf 0.62$ & $\bf 3.6$ & $\bf 46.4$ &$\bf 43.4$ &$\bf 1.8$ &  $\bf 1.4$ \\

& & & & & & & & &  \\
 Run1$_{a}$& 3 & 0 & 57.3 &   0.67 & 5 & 48 & 36.2  & 3.3 & 3.2 \\
 Run1$_{b}$ & 3 & 0 & 64.8 &   0.58 & 3 & 41 & 21.2  & 1.5 & 12.5  \\
 Run1$_{c}$ & 2 & 1 & 47.5 &   0.58 & 4 & 35 & 42.7 & 2.3 & 7.5  \\ 
 Run1$_{d}$ & 2 & 0 & 50.6 &   0.71 & 5 & 45 & 44.6 & 2.5 & 2.3  \\ 
 Run1$_{e}$ & 2 & 0 & 42.5 &   0.56 & 5 & 45 & 43.3 & 6.2 & 8.0  \\  
 $\bf Run1_{ave}$ & $\bf 2.4$ & $\bf 0.2$ & $\bf 52.5$ & $\bf 0.62$ & $\bf 4.4$ & $\bf 42.8$ & $\bf 37.6$ & $\bf 3.2$ & $\bf 6.7$ \\  
 
 & & & & & & & & &  \\
 Run10$_{a}$ & 3 & 0 & 55.6 &   0.71 & 2 & 15 & 14.8  & 1.7 & 27.9  \\
 Run10$_{b}$ & 3 & 0 & 50.8 &   0.63 & 4 & 7 & 26.2  & 1.5 & 21.5   \\
 Run10$_{c}$ & 4 & 0 & 62.1 &   0.59 & 5 & 2 & 10.6  & 1.3 & 26.0  \\
 Run10$_{d}$ & 3 & 0 & 56.5 &   0.55 & 3 & 2 & 14.6 & 1.5 & 27.4   \\  
 Run10$_{e}$ & 2 & 0 & 45.4 &   0.52 & 2 & 2 & 24.6 & 3.7 & 26.3  \\  
 $\bf Run10_{ave}$   & $\bf 3.0$ & $\bf 0.0$ & $\bf 54.1$ & $\bf 0.60$ & $\bf 3.2$ & $\bf 5.6$ & $\bf 18.2$ & $\bf 1.9$ & $\bf 25.8$ \\
 
  & & & & & & & & &  \\
 RunJ0$_{a}$ & 2 & 0 & 54.6 & 0.69 & 2 & 0 & 6.9 & 3.3 & 35.2 \\
 RunJ0$_{b}$ & 3 & 0 & 60.4 & 0.54 & 1 & 0 & 2.3 & 1.3 & 35.9 \\
 RunJ0$_{c}$ & 3 & 0 & 48.5 & 0.45 & 1 & 1 & 9.8 & 6.3 & 35.4 \\
 RunJ0$_{d}$ & 4 & 0 & 56.9 & 0.65 & 1 & 1 & 2.3 & 4.4 & 36.3 \\
 RunJ0$_{e}$ & 2 & 0 & 38.8 & 0.65 & 1 & 0 & 19.4 & 2.9 & 38.8 \\
 RunJ0$_{f}$ & 2 & 0 & 52.7 & 0.65 & 1 & 0 & 7.1  & 3.5 & 36.7 \\
 $\bf RunJ0_{ave}$  & $\bf 2.7$ & $\bf 0.0 $& $\bf 52.0$ & $\bf 0.61$ &$\bf 1.2$&$\bf 0.3$&$\bf 8.0$ & $\bf 3.6$ & $\bf 36.4$ \\ 

 & & & & & & & & &  \\
 RunJ$_{a}$ & 3 & 0 & 47.3 & 0.39 & 1 & 0 & 2.3 & 10.2 & 40.2 \\
 RunJ$_{b}$ & 3 & 0 & 49.4 & 0.59 & 0 & 0 & 0.0 & 8.8 & 41.7 \\
 RunJ$_{c}$ & 2 & 0 & 48.1 & 0.57 & 1 & 0 & 6.7 & 10.6 & 34.6 \\
 RunJ$_{d}$ & 3 & 0 & 48.5 & 0.31 & 2 & 0 & 8.5 & 6.0 & 37.1 \\
 RunJ$_{e}$ & 3 & 0 & 44.2 & 0.55 & 1 & 0 & 1.9 & 7.7 & 46.1 \\
 RunJ$_{f}$ & 4 & 0 & 51.9 & 0.63 & 1 & 0 & 4.4 & 8.3 & 35.4 \\
 $\bf RunJ_{ave}$   & $\bf 3.0$ & $\bf 0.0$ & $\bf 48.2$ & $\bf 0.51$ & $\bf 1.0$ & $\bf 0.0$ & $\bf 4.0$ & $\bf 8.6$ & $\bf 39.2$ \\
 
   & & & & & & & & &  \\
 RunJS$_{a}$ & 3 & 0 & 46.5 &   0.55 & 1 & 0 &  1.9   & 14.5 & 37.1   \\ 
 RunJS$_{b}$ & 2 & 0 & 42.3 &   0.50 & 1 & 1 &  2.3  & 21.1 & 34.3   \\
 RunJS$_{c}$ & 3 & 0 & 44.8 &   0.46 & 0 & 1 &  0.2  & 24.4 & 30.6  \\  
 RunJS$_{d}$ & 3 & 0 & 45.8 &   0.35 & 1 & 0 &  1.9  & 21.9 & 30.4   \\ 
 RunJS$_{e}$ & 4 & 0 & 41.2 &   0.35 & 1 & 2 &  2.3  & 21.1 & 35.4   \\
 RunJS$_{f}$ & 3 & 0 & 44.8 &   0.34 & 0 & 0 &  0.0  & 21.3 & 33.9  \\ 
 $\bf RunJS_{ave}$  & $\bf 3.0$ & $\bf 0.0$ & $\bf  44.2$ & $\bf 0.43$ & $\bf 0.7 $ & $\bf 0.7$ &$\bf 1.4$ & $\bf 20.7$ & $\bf 33.6$ \\

\enddata 
\end{deluxetable} 

%%%%%%%%%%%%%%%%%%%%%%%%%%%%%%%%%%%%%%%%%
\clearpage
\begin{deluxetable}{llllllllllll}
%\rotate
\tablecolumns{12} 
%\tabletypesize{\scriptsize} 
\tabletypesize{\footnotesize} 
\tablecaption{Final Planets with $M_f  >$ 0.5 \mearth and  0.72 AU $< a_f < $ 1.52 AU
\label{tbl-3}} 
\tablewidth{0pt} 
\tablehead{
\colhead{$ \bf  Run $}       &
\colhead{$ \bf Planet $}   &
\colhead{$ \bf \massf $}  &
\colhead{$ \bf a $}  & 
\colhead{$ \bf e $}     &
\colhead{$ \bf i $}     &
\colhead{$ \bf \mwater$} & 
\colhead{$ \bf \logwmf$} & 
\colhead{$ \bf \mwater$} & 
\colhead{$ \bf \logwmf$} &
\colhead{$ \bf t_{EM} $}   &
\colhead{$ \bf t_{PL} $} \\
\colhead{} &
\colhead{} &
\colhead{} &
\colhead{} &
\colhead{} &
\colhead{} &
\colhead{Step } &
\colhead{Step } &
\colhead{Power-law } &
\colhead{Power-law } &
\colhead{} &
\colhead{} \\
\colhead{} & 
\colhead{} & 
\colhead{(\mearth)} & 
\colhead{(AU)} & 
\colhead{} & 
\colhead{($^\circ$)} & 
\colhead{(\mocean)} &
\colhead{} & 
\colhead{(\mocean)} & 
\colhead{} & 
\colhead{(Myr)} & 
\colhead{(Myr)} 
}
\startdata 
%\nodata & Earth  & 1.000 & 1.000 & 0.02 & 0 & $\sim$10 & $\sim$-2.6 &  $\sim$10 & $\sim$-2.6 & \nodata & \nodata \\
\nodata & Earth  & 1.000 & 1.000 & 0.02 & 0 & $\sim$5 & $\sim$-2.9 &  $\sim$5 & $\sim$-2.9 & \nodata & \nodata \\
&&&&&&&&&&& \\
 Run0$_a$ & EM06 & 1.0640 & 1.1367 & 0.08 & 2.55 & 11.8317 & -2.5560 & 15.8749 & -2.4283 & 89.8 &  948.1 \\ 
 Run0$_b$ & EM04 & 0.9707 & 0.9861 & 0.06 & 6.03 & 9.8877 & -2.5940 & 16.5639 & -2.3700 & 587.4 & 638.8  \\ 
 Run0$_c$ & EM03 & 0.9893 & 0.8139 & 0.02 & 5.83 & 13.2884 & -2.4739 & 14.3262 & -2.4413 & 42.1  & 931.6  \\ 
 Run0$_c$ & EM06 & 1.4373 & 1.5051 & 0.12 & 1.42 & 92.0972 & -1.7954 & 91.2174 & -1.7995 & 845.3 & 468.5  \\ 
 Run0$_d$ & EM02 & 1.1853 & 0.7614 & 0.07 & 4.32 & 31.8483 & -2.1728 & 23.5678 & -2.3036 & 94.1 &  574.8  \\ 
 Run0$_d$ & EM08 & 0.5787 & 1.2632 & 0.07 & 0.08 & 11.2948 & -2.3116 & 12.5914 & -2.2644 & 7.1 & 948.7  \\ 
 Run0$_e$ & EM03 & 1.3813 & 0.7960 & 0.03 & 7.15 & 24.5758 & -2.3519 & 28.5346 & -2.2870 & 90.7 & 683.5  \\ 
 $\bf Run0_{ave}$  & \nodata & \nodata & \nodata & \nodata & \nodata &27.8320 & \nodata &27.9198& \nodata & \nodata & \nodata  \\

&&&&&&&&&&& \\
 Run1$_a$ & EM08 & 1.3533 & 1.1613 & 0.07 & 1.95 & 60.8474 & -1.9492 & 55.9253 & -1.9859 & 446.4 &  350.4 \\ 
 Run1$_b$ & EM13 & 0.8400 & 0.9964 & 0.01 & 6.49 & 13.2455 & -2.4043 & 13.7572 & -2.3878 & 9.3 & 500.0  \\ 
 Run1$_c$ & EM08 & 1.1293 & 1.2388 & 0.04 & 2.97 & 37.5928 & -2.0798 & 42.3437 & -2.0281 & 146.9  & 720.2  \\ 
 Run1$_d$ & EM11 & 1.2787 & 1.3307 & 0.16 & 4.67 & 128.0309 & -1.6015 & 126.4799 & -1.6068 &847.0  & 628.6  \\ 
 Run1$_e$ & EM10 & 0.8680 & 1.2351 & 0.07 & 3.24 & 39.6704 & -1.9421 & 24.0501 & -2.1595 & 22.6 & 566.5  \\ 
 $\bf Run1_{ave}$  & \nodata & \nodata & \nodata & \nodata & \nodata &55.8774& \nodata &50.5728& \nodata & \nodata & \nodata  \\

&&&&&&&&&&& \\
 Run10$_a$ & EM03 & 0.9707 & 1.0072 & 0.08 & 4.14 & 22.6931 & -2.2332 & 21.1531 & -2.2638 & 139.0 & 101.7  \\ 
 Run10$_b$ & EM02 & 1.1480 & 0.7481 & 0.07 & 6.07 & 9.9318 & -2.6650 & 13.9879 & -2.5163 & 41.1 & 295.3  \\ 
 Run10$_c$ & EM07 & 0.9800 & 1.3014 & 0.11 & 1.80 & 47.3618 & -1.9179 & 42.5852 & -1.9640 & 52.9 & 195.9  \\ 
 Run10$_c$ & EM08 & 0.9613 & 0.7796 & 0.09 & 4.22 & 5.7113 & -2.8282 & 7.1824 & -2.7287 & 18.2 & 137.2  \\ 
 Run10$_d$ & EM03 & 0.9427 & 0.8229 & 0.06 & 3.86 & 7.6507 & -2.6927 & 11.9026 & -2.5008 & 25.2 & 206.6  \\ 
 Run10$_e$ & EM05 & 1.7827 & 1.0853 & 0.08 & 4.02 & 34.4407 & -2.3161 & 37.6937 & -2.2769 & 501.3 & 76.9  \\ 
 $\bf Run10_{ave}$  & \nodata & \nodata & \nodata & \nodata & \nodata &21.2982& \nodata &23.6404& \nodata & \nodata & \nodata  \\

 &&&&&&&&&&& \\
 RunJ0$_b$ & EM08 & 1.7360 & 1.0839 & 0.0721 & 3.2170 & 32.0182 & -2.3362 & 29.8345 & -2.3669  & 142.6&119.3 \\
 RunJ0$_c$ & EM04 & 0.9427 & 0.9897 & 0.0285 & 6.7830 & 9.8866 & -2.5814 & 14.3867 & -2.4185  &28.6 &214.4 \\
 RunJ0$_d$ & EM02 & 0.6720 & 0.7468 & 0.0851 & 9.0870 & 26.2659 & -2.0100 & 23.6281 & -2.0560&21.3 & 331.4\\
 RunJ0$_d$ & EM04 & 1.0267 & 1.1357 & 0.0523 & 5.6770 & 10.0748 & -2.6103 & 13.7350 & -2.4757&19.4 &106.9 \\
 RunJ0$_e$ & EM02 & 1.3253 & 0.8828 & 0.0832 & 3.3210 & 48.7984 & -2.0360 & 27.8365 & -2.2798&81.2 &59.3 \\
 RunJ0$_f$ & EM03 & 0.9333 & 1.4802 & 0.1014 & 4.6340 & 47.5447 & -1.8950 & 51.8762 & -1.8571& 55.9&146.5 \\
 RunJ0$_f$ & EM04 & 1.6240 & 0.7384 & 0.1487 & 0.8150 & 10.2095 & -2.8036 & 19.5181 & -2.5222  & 186.0& 84.9\\ 
 $\bf RunJ0_{ave}$  & \nodata & \nodata & \nodata & \nodata & \nodata &26.3997& \nodata &28.4658& \nodata & \nodata & \nodata  \\

 &&&&&&&&&&& \\
 RunJ$_a$ & EM16 & 1.0360 & 1.5068 & 0.0690 & 0.7780 & 20.1828 & -2.3124 & 26.9599 & -2.1867  & 31.2& 39.5\\
 RunJ$_b$ & EM03 & 1.0360 & 1.0336 & 0.0181 & 8.7380 & 22.9914 & -2.2559 & 15.9460 & -2.4148  & 413.8&95.1 \\ 
 RunJ$_c$ & EM06 & 1.4840 & 1.2891 & 0.1753 & 5.7690 & 21.3647 & -2.4438 & 23.4678 & -2.4030  &170.2 &38.3 \\
 RunJ$_d$ & EM03 & 0.5787 & 0.9071 & 0.0336 & 1.5680 & 0.0971 & -4.3774 & 2.0838 & -3.0456  & 2.4&51.5 \\ 
 RunJ$_d$ & EM10 & 1.0920 & 1.3786 & 0.0384 & 2.3500 & 21.0534 & -2.3170 & 19.7243 & -2.3453  &24.2 &175.4 \\ 
 RunJ$_e$ & EM03 & 0.9053 & 1.0559 & 0.0589 & 3.2300 & 0.1471 & -4.3913 & 4.1579 & -2.9400  & 51.8&86.0 \\ 
 RunJ$_f$ & EM03 & 1.1573 & 0.7322 & 0.0572 & 6.0330 & 18.8940 & -2.3892 & 10.2374 & -2.6553  & 73.0&226.9 \\ 
 $\bf RunJ_{ave}$  & \nodata & \nodata & \nodata & \nodata & \nodata &14.9615& \nodata &14.3710& \nodata & \nodata & \nodata  \\

 &&&&&&&&&&& \\
 RunJS$_a$ & EM05 & 1.1107 & 0.8915 & 0.06 & 0.34 & 2.7238 & -3.2125 & 11.7314 & -2.5783 & 51.9 & 40.4  \\ 
 RunJS$_b$ & EM10 & 0.8960 & 1.3564 & 0.14 & 2.79 & 0.4794 & -3.8737 & 8.5170 & -2.6241 & 37.1 & 96.1  \\ 
 RunJS$_c$ & EM04 & 0.8867 & 0.9807 & 0.08 & 5.08 & 39.2646 & -1.9558 & 18.7532 & -2.2767 & 172.9 & 171.3  \\ 
 RunJS$_c$ & EM08 & 0.6627 & 1.4902 & 0.11 & 3.14 & 0.1004 & -4.4215 & 3.8900 & -2.8334 & 30.7 & 32.8  \\ 
 RunJS$_d$ & EM05 & 0.8960 & 1.4412 & 0.03 & 4.99 & 4.5815 & -2.8934 & 10.8548 & -2.5187 & 13.3 & 64.8  \\ 
 RunJS$_d$ & EM11 & 0.7840 & 0.9106 & 0.10 & 0.87 & 0.1792 & -4.2430 & 5.5202 & -2.7544 & 27.8 & 355.2  \\ 
 RunJS$_e$ & EM03 & 0.9427 & 0.7578 & 0.02 & 4.18 & 0.1116 & -4.5287 & 2.3959 & -3.1969 & 10.6 & 119.3  \\ 
 RunJS$_e$ & EM09 & 0.5880 & 1.3696 & 0.03 & 3.49 & 18.7234 & -2.0991 & 25.6518 & -1.9623 & 48.4 & 105.0  \\ 
 RunJS$_f$ & EM08 & 1.0080 & 0.8699 & 0.16 & 5.88 & 1.9805 & -3.3087 & 5.0155 & -2.9052 & 22.1  & 60.6  \\ 
 $\bf RunJS_{ave}$  & \nodata & \nodata & \nodata & \nodata & \nodata &7.5716& \nodata &7.3160& \nodata & \nodata & \nodata  \\

\enddata 
%\tablenotetext{$\ddagger$}{Power-law disk \wmf distribution.}

\end{deluxetable} 
\clearpage
%%%%%%%%%%%%%%%%%%%%%%%%%%%%%%%%%%%%%%%%%

\clearpage

\begin{deluxetable}{lllll}
\tablecolumns{5} 
\tabletypesize{\footnotesize} 
\tablecaption{Final Planets and the Embryos/Planetesimals Each Accreted
\label{tbl-2}} 
\tablewidth{0pt} 
\tablehead{
\colhead{$\bf Run$} &
\colhead{$\bf Planet$} &
\colhead{$\bf Component$}   &
\colhead{$\bf M \; (\mearth)$} &
\colhead{$\bf a \; (AU)$}}
\startdata 

 RunJS$_e$ & EM01 &  \nodata & 0.2147 & 0.4453  \\  
 \nodata & \nodata & EM01 & 0.0933 & 0.4590 \\  
 \nodata & \nodata & PL001 & 0.0093 & 0.3570 \\  
 \nodata & \nodata & PL002 & 0.0093 & 0.4340 \\  
 \nodata & \nodata & PL003 & 0.0093 & 0.4686 \\  
 \nodata & \nodata & PL005 & 0.0093 & 0.5176 \\  
 \nodata & \nodata & PL006 & 0.0093 & 0.5370 \\  
 \nodata & \nodata & PL007 & 0.0093 & 0.5543 \\  
 \nodata & \nodata & PL008 & 0.0093 & 0.5701 \\  
 \nodata & \nodata & PL031 & 0.0093 & 0.7942 \\  
 \nodata & \nodata & PL033 & 0.0093 & 0.8119 \\  
 \nodata & \nodata & PL036 & 0.0093 & 0.8389 \\  
 \nodata & \nodata & PL041 & 0.0093 & 0.8849 \\  
 \nodata & \nodata & PL056 & 0.0093 & 1.0301 \\  
 \nodata & \nodata & PL122 & 0.0093 & 1.8002 \\  

 RunJS$_e$ & EM11 &  \nodata & 0.2520 & 0.9866  \\  
 \nodata & \nodata & EM08 & 0.0933 & 1.2482 \\  
 \nodata & \nodata & EM11 & 0.0933 & 1.6021 \\  
 \nodata & \nodata & PL059 & 0.0093 & 1.0605 \\  
 \nodata & \nodata & PL095 & 0.0093 & 1.4594 \\  
 \nodata & \nodata & PL097 & 0.0093 & 1.4834 \\  
 \nodata & \nodata & PL105 & 0.0093 & 1.5814 \\  
 \nodata & \nodata & PL107 & 0.0093 & 1.6064 \\  
 \nodata & \nodata & PL129 & 0.0093 & 1.8944 \\  
 \nodata & \nodata & PL150 & 0.0093 & 2.1831 \\  
 RunJS$_e$ & EM16 &  \nodata & 0.0933 & 2.4220  \\  
 \nodata & \nodata & EM16 & 0.0933 & 2.2899 \\  
 RunJS$_e$ & PL198 &  \nodata & 0.0093 & 3.1238  \\  
 \nodata & \nodata & PL198 & 0.0093 & 2.9433 \\  
 RunJS$_e$ & PL248 &  \nodata & 0.0093 & 18.1745  \\  
 \nodata & \nodata & PL248 & 0.0093 & 3.8552 \\  
 \enddata 
\tablecomments{For each simulation (noted in Column 1), each of the final planets is listed (in Column 2) along with all embryos and/or planetesimals that compose that planet (given in Column 3).  The masses (Column 4) and semimajor axes (Column 5) are given for each initial body.  Because the initial water distribution in the disk is assigned by specifying the water mass fraction for each embryo/planetesimal according to the distance from the central star, this table supplies all of the necessary information needed to examine the effects of alternative water distributions.  This table is published in its entirety in the electronic edition of the Astrophysical Journal. A portion is shown here for guidance regarding its form and content.}
\end{deluxetable} 
%%%%%%%%%%%%%%%%%%%%%%%%%%%%%%%%%%%%%%%%%

%%%%%%%%%%%%%%%%%%%%%%%%%%%%%%%%%%%%%%%%%
\clearpage

\begin{figure*} % FIGURE 1
\includegraphics[width=.95\textwidth]{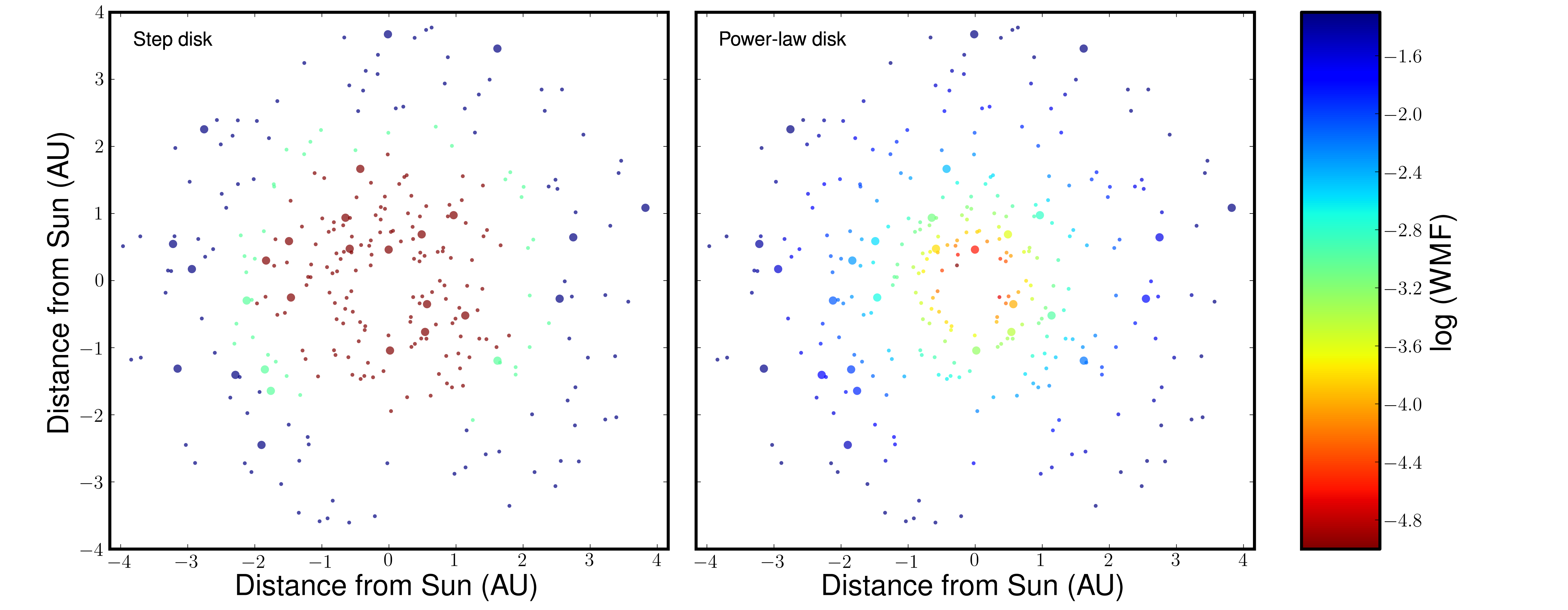}
\figcaption{View of the initial disk used in our simulations projected into the disk midplane.  Sizes of the bodies are enhanced by a factor of more than 10$^4$ relative to the distance scale.  Colors in the left panel correspond to the distribution of \water used by \citet{Raymond.etal:2004}, whereas those in the panel on the right show a disk with the same total mass of \water but distributed using a smooth power-law function.}
\end{figure*}

\begin{figure*} % FIGURE 2
\includegraphics[width=.8\textwidth]{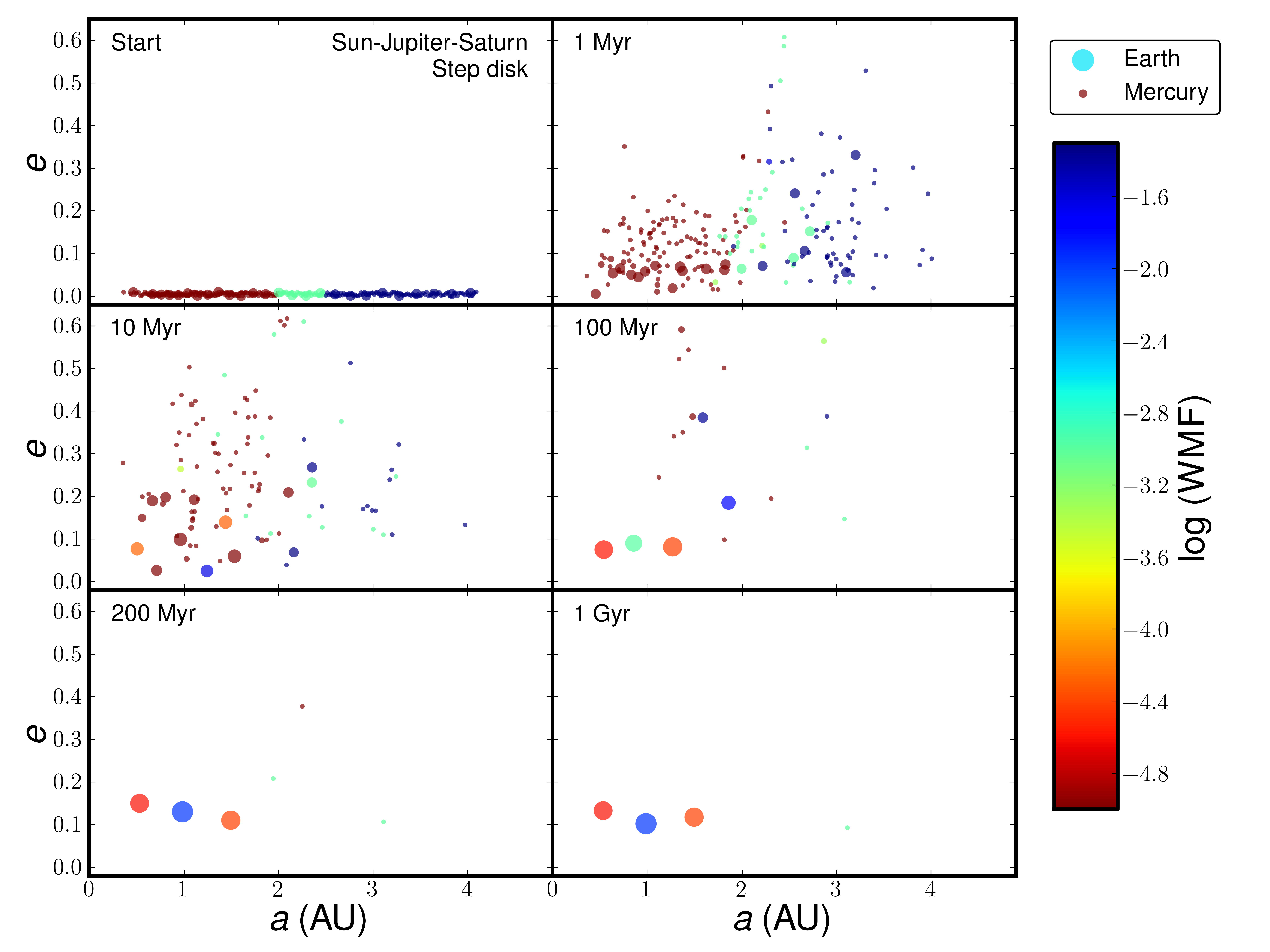}
\figcaption{Evolution of our step disk model around the Sun with Jupiter and Saturn perturbing the system.  Both outer planets began on orbits with $e$ = 0.05.  Each panel shows the eccentricity versus semimajor axis for each body in the disk that orbits interior to Jupiter at the time noted.  The size of each symbol is proportional to that of each body.  The colors represent the water mass fraction, where red bodies are dry and blue bodies are water rich.  The legend provides symbols of an Earth-size planet with 10 oceans of water (our definition for a `water-rich' planet) along with a dry Mercury for comparison with the final planets.  In this simulation, a 0.9 \mearth planet formed at 1 AU with 39 oceans of water, whereas the two adjacent planets remained dry, accreting less than 0.1 \mocean each. One dry planetesimal that did not accrete any material remained in the system at 3.1 AU.}
\end{figure*}

\begin{figure*} % FIGURE 3
\includegraphics[width=.8\textwidth]{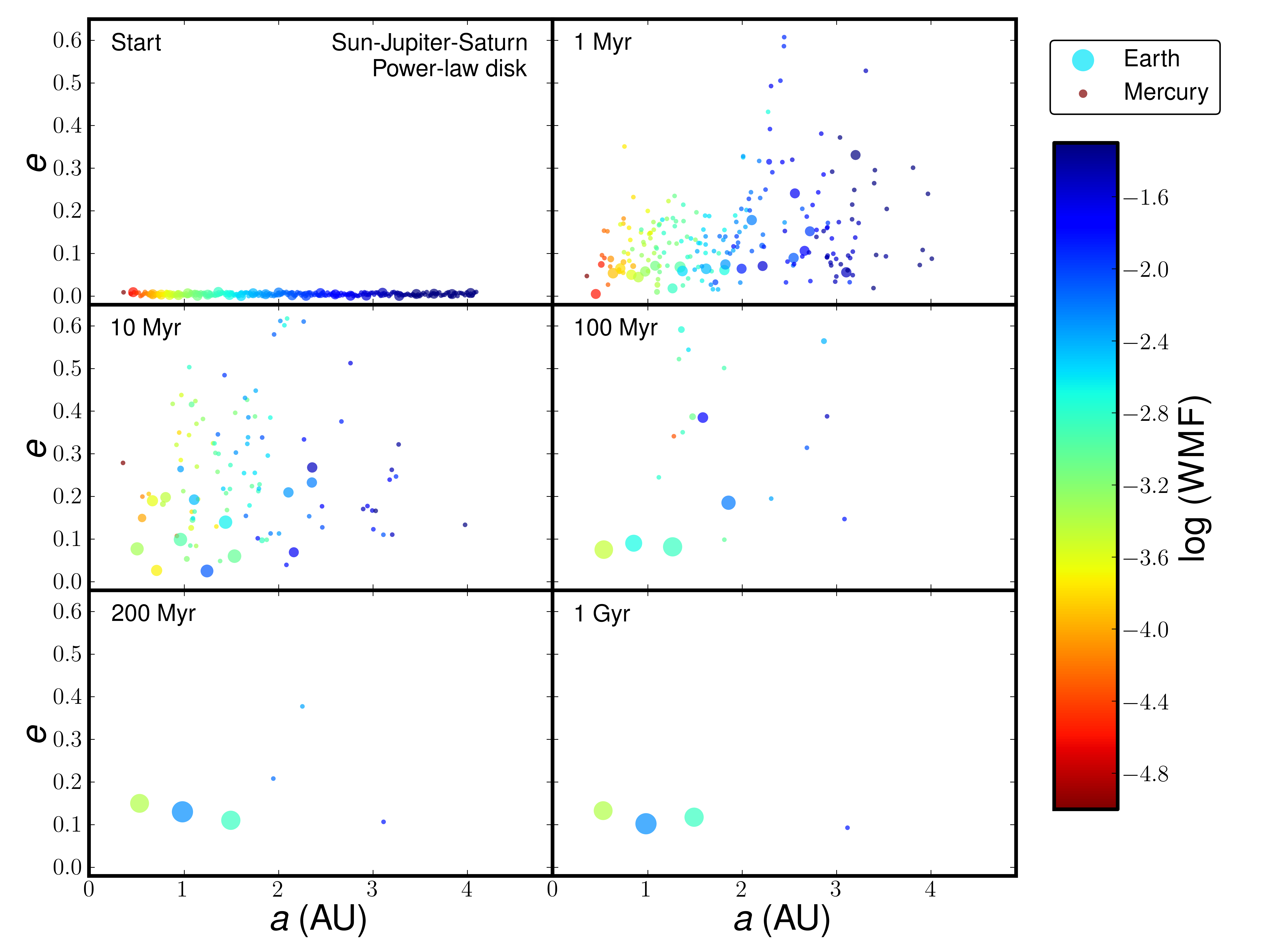}
\figcaption{Evolution of our smooth power-law disk model for the same simulation that is shown in Figure 2.  The symbols and color scheme are described in the Figure 2 caption. In this case, the inner planet remained dry, the Earth-analog near 1 AU remained water-rich (with 19 oceans), and the planet near 1.5 AU accreted a moderate amount (4 \mocean) of water.}
\end{figure*}

\begin{figure*} % FIGURE 4
\includegraphics[width=.8\textwidth]{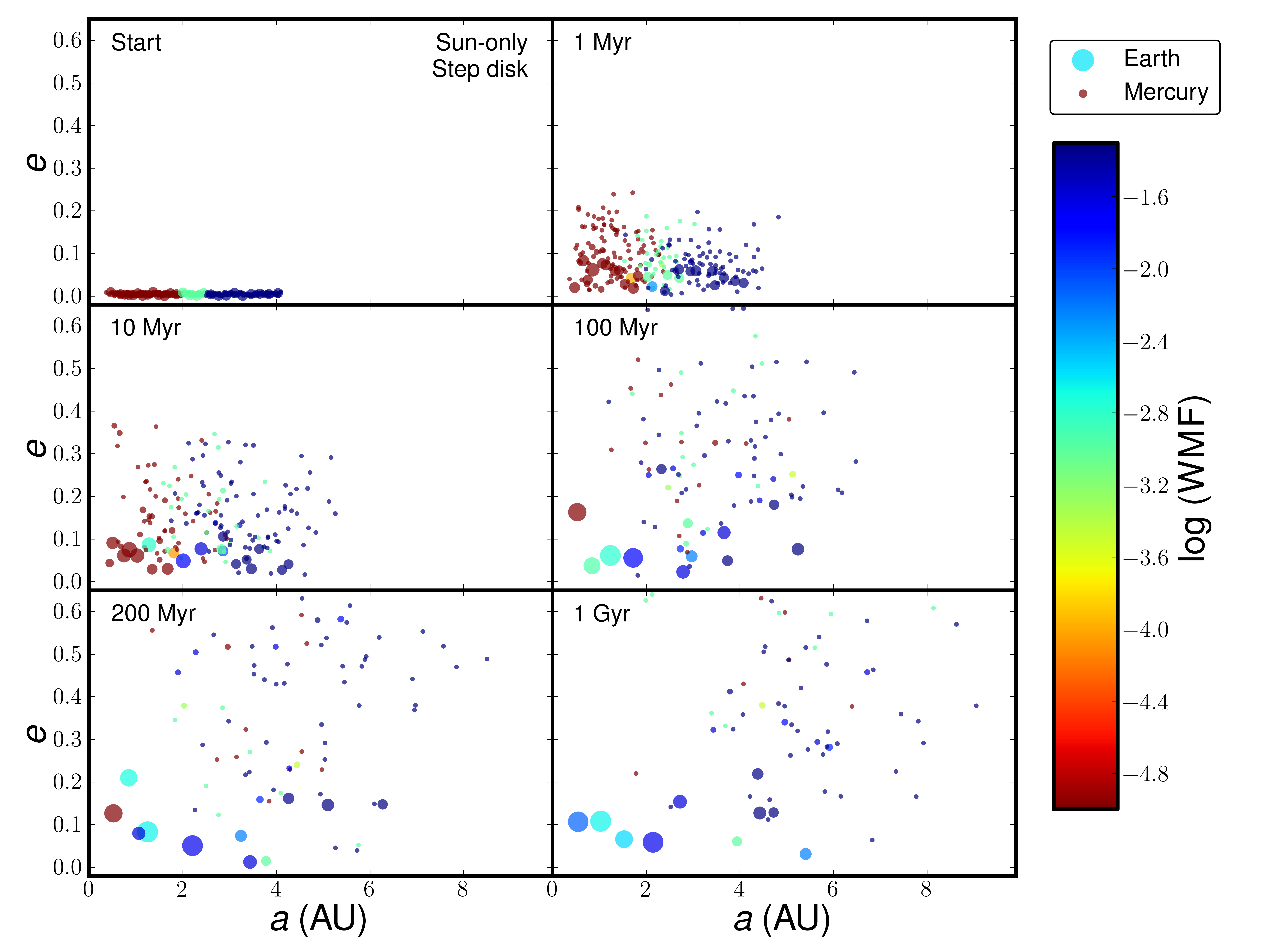}
\figcaption{Evolution of our step disk model around the Sun without any outer planets perturbing the system.  The symbols and color scheme are described in the Figure 2 caption.  Note the expanded range of the horizontal scale.  In this simulation, eight planets formed between 0.5 and 7 AU, with an additional 47 planetesimals still in the system between 3 and 28 AU.  Five of the planets accreted more than 10 oceans and the other three planets were moderately wet, accreting at least 4 oceans of water. Even without giant planets perturbing the system, all eight planets were able to accrete volatile-rich material from beyond the snow-line.}
\end{figure*}

\begin{figure*} % FIGURE 5
\includegraphics[width=.8\textwidth]{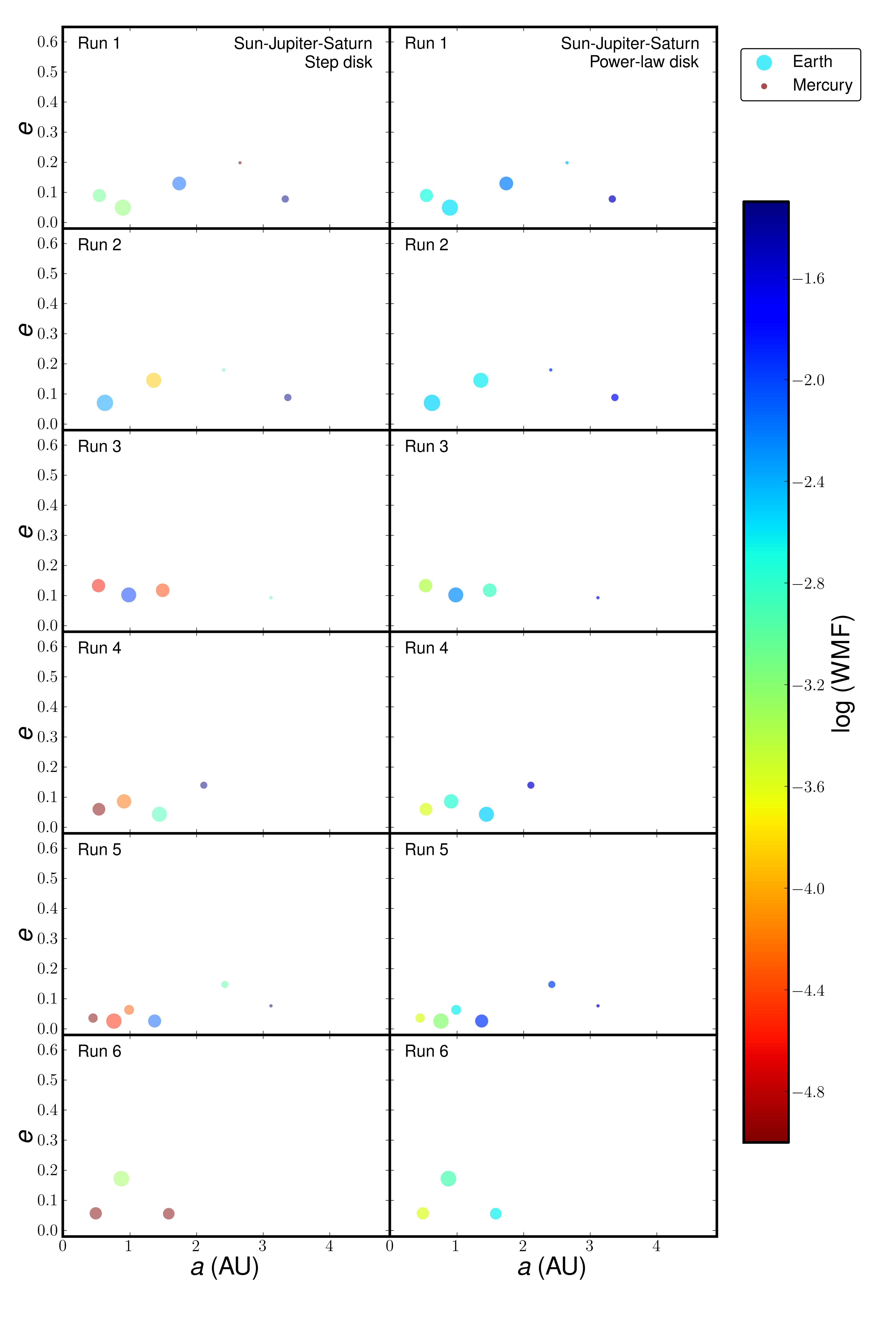}
\figcaption{Final planets that formed in each of the six simulations with Jupiter and Saturn perturbing the system. Each panel shows the eccentricities and semimajor axes for each body at the end of a given simulation. The left column shows water abundances assuming an initial step disk distribution, and the right column shows the results from the same six simulations but using our smooth power-law disk model.   The symbols and color scheme are described in the Figure 2 caption.  Each final system of 3 -- 5 planets (and 0 -- 2 minor planets) was composed of $\sim$45\% of the initial disk mass, and the remaining mass was either perturbed into the Sun (20\% on average) or was ejected from the system ($\sim$34\%).  The final distribution of water in these planets varied among the simulations as well as between the different initial disks used.  From 0 -- 2 planets were water-rich ($>$10\mocean by our definition) in the step disk runs and from 0 -- 3 were wet in the power-law disk runs, but on average each disk produced the same number of wet planets.  More dry planets ($<$2 \mocean), however, formed in the step disk runs than in the power-law disk runs, and only the step disk runs produced more dry planets than water-rich (or even moderately wet) planets.}
\end{figure*}

\begin{figure*} % FIGURE 6
\includegraphics[width=.8\textwidth]{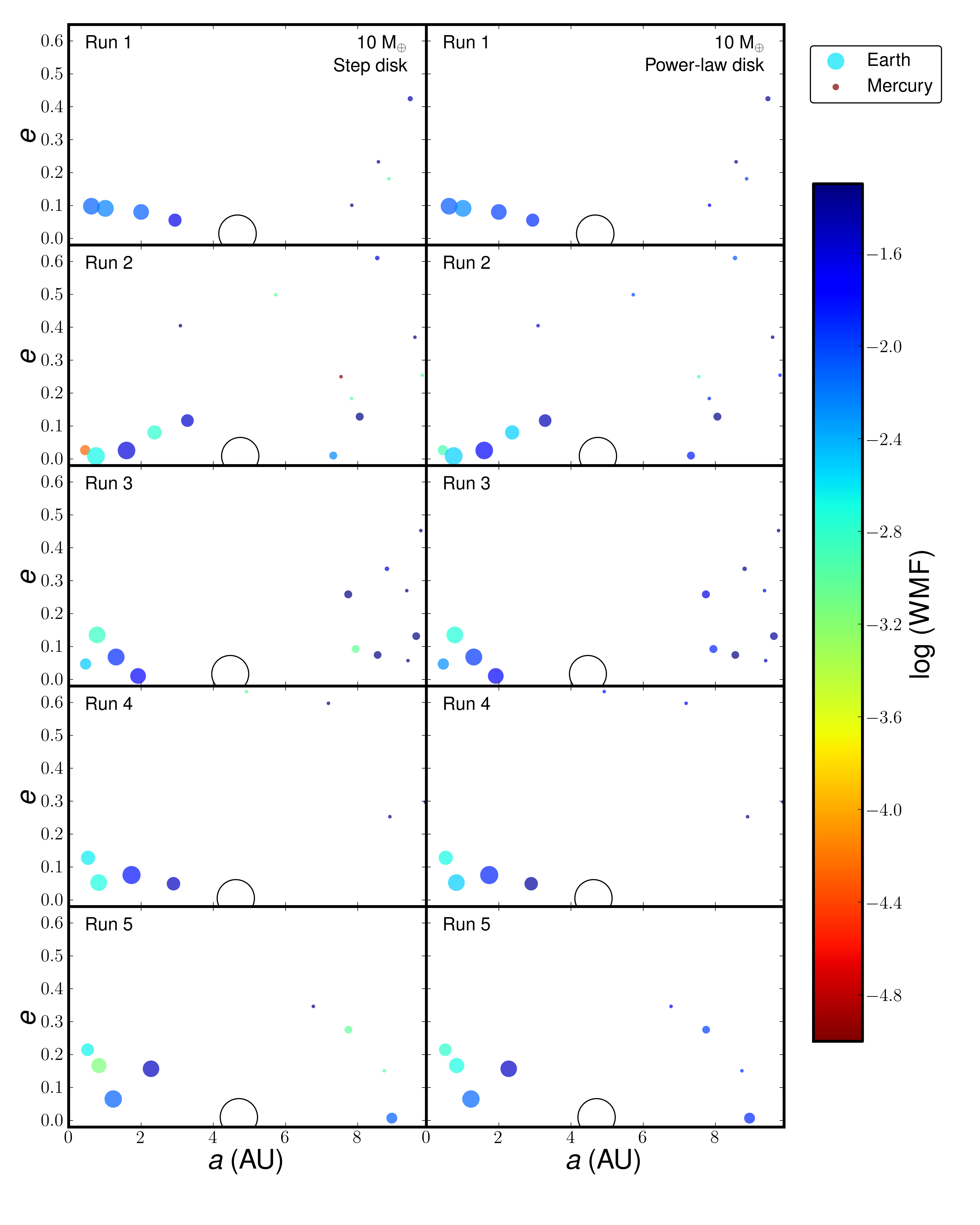}
\figcaption{Final planets that formed in simulations with a 10 \mearth companion in Jupiter's orbit (5.2 AU). The symbols and color scheme are described in the Figure 2 caption, and the panels are in the same format as described in the Figure 5 caption. Note the expanded range of the horizontal scale as compared to Figure 5.  The open black circle represents the 10 \mearth planet.  From 4 -- 9 planets formed, and up to 15 minor planets remained, in each simulation. More than 70\% of the initial disk mass was accreted, and most of the remaining mass was ejected from the system.  Note that bound planets beyond 10 AU do not appear in the figure but contribute to these totals.   Unlike the simulations with Jupiter and Saturn, most of the planets were water-rich or moderately wet (2 - 10 \mocean by our definition), and only 1 or 2 dry planets remained in two of the systems.  Differences in water distribution in the final planets formed from both disk models were not significant.}
\end{figure*}

\begin{figure*} % FIGURE 7
\includegraphics[width=.8\textwidth]{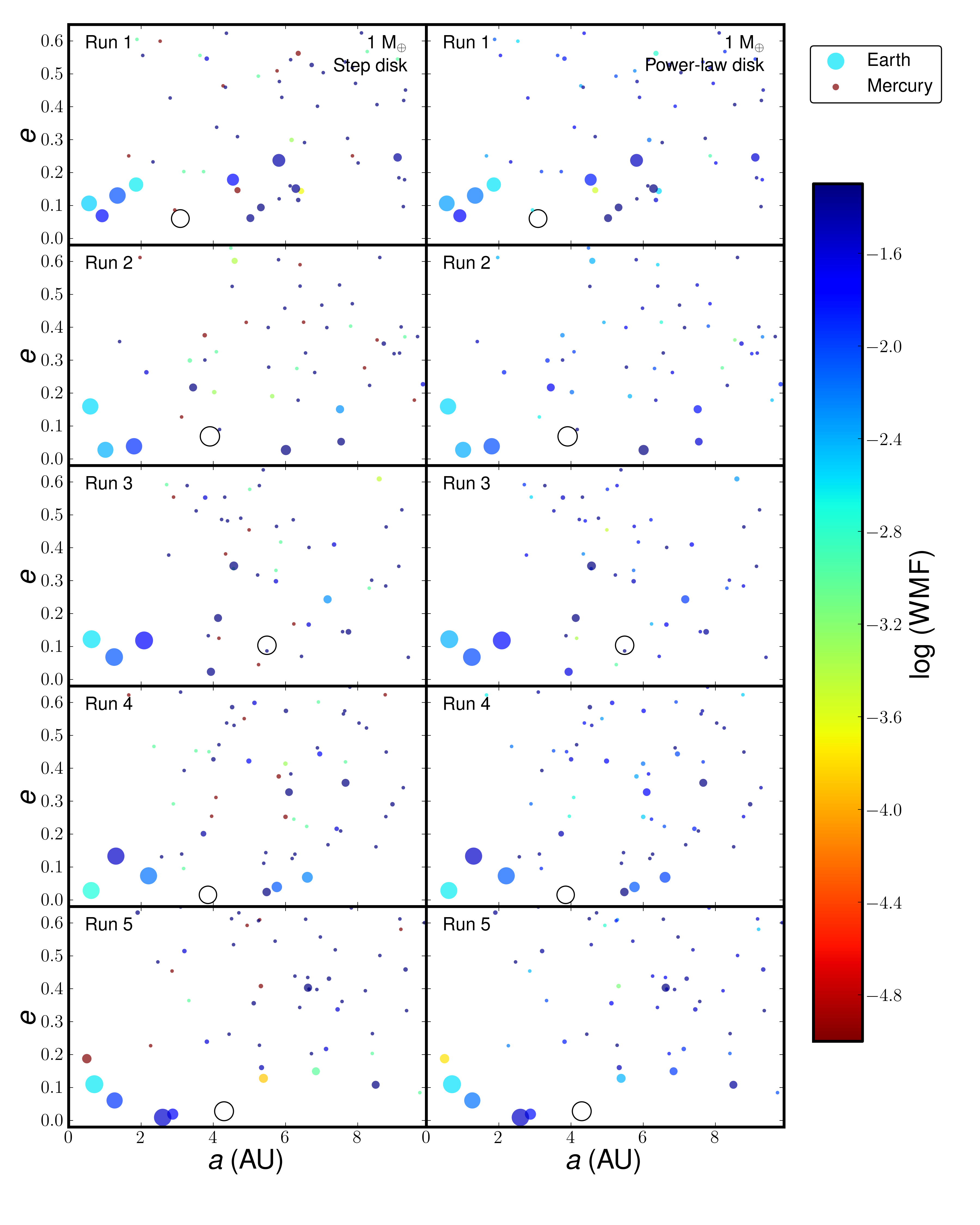}
\figcaption{Final planets that formed in simulations with a 1 \mearth companion in Jupiter's orbit (5.2 AU).  The symbols and color scheme are described in the Figure 2 caption, and the panels are in the same format as described in the Figure 5 caption. Note the expanded range of the horizontal scale as compared to Figure 5.  The open black circle represents the 1 \mearth planet that began at 5.2 AU.  In these simulations, 6 -- 8 planets (and 35 -- 48 minor planets) formed with more than 90\% of the initial disk mass.  Most of the planets (4 -- 7) were water-rich, 0 -- 2 were moderately wet, and only one simulation formed dry planets.  The number of wet, moderately wet, and dry planets was virtually the same between the two disk models for all but one simulation.}
\end{figure*}

\begin{figure*} % FIGURE 8
\includegraphics[width=.8\textwidth]{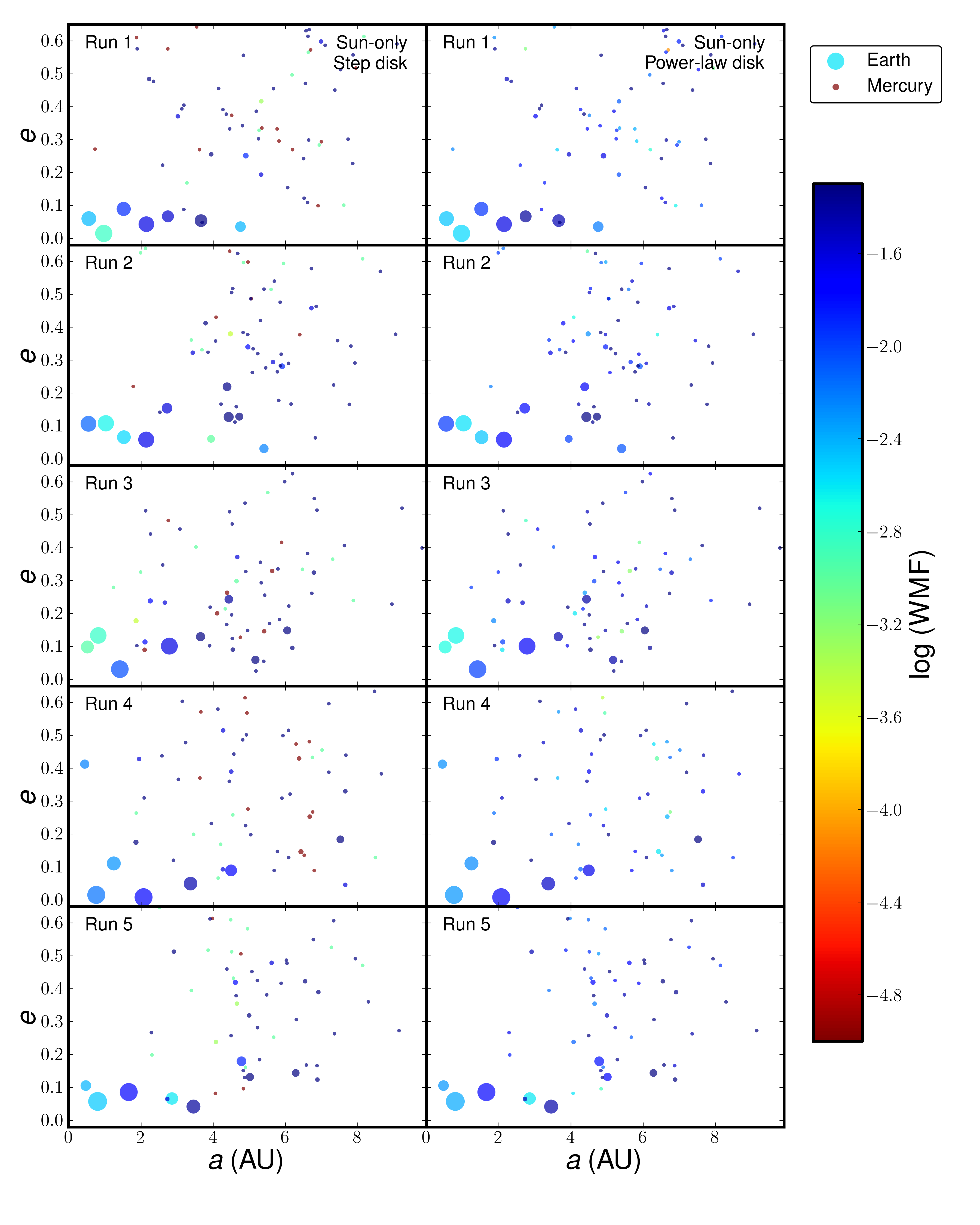}
\figcaption{Final planets that formed in simulations with no outer planet companion.   The symbols and color scheme are described in the Figure 2 caption, and the panels are in the same format as described in the Figure 5 caption. Note the expanded range of the horizontal scale as compared to Figure 5.   From 5 -- 8 planets formed, and a larger number of minor planets (38 -- 52) remained, in each simulation, and only 3\% (on average) of the initial disk mass was lost.  From 5 -- 6 of the planets were water-rich, 0 -- 3 were moderately wet, and no simulation formed a dry planet.  Varying the initial water distribution led to virtually the same results in terms of the number of wet, and moderately wet, planets that formed.}
\end{figure*}

\bibliographystyle{apj}
\bibliography{references2}

\end{document}